\documentclass[aps,prb]{revtex4}
\usepackage{bm}
\usepackage{epsfig}

\def\bm{\bf}

\begin{document}
\title{D'yakonov-Perel' spin relaxation under electron-electron collisions in
{\it n}-type QWs}
\author{M.M.~Glazov\dag, E.L.~Ivchenko\dag, M.A.~Brand\ddag,
O.Z.~Karimov\ddag $\mbox{}$ and R.T.~Harley\ddag}
\affiliation{\dag Ioffe Physical-Technical Institute,
Russian Academy of Sciences 194021 St. Petersburg, Russia\\
\ddag Department of Physics and Astronomy, University of
Southampton Southampton SO17 1BJ, United Kingdom}

\begin{abstract} The D'yakonov-Perel' spin relaxation mechanism in {\it
n}-doped GaAs/AlGaAs quantum wells (QWs) has been studied both
theoretically and experimentally. The temperature dependence of
the spin relaxation time has been calculated for arbitrary
degeneracy of the 2D electron gas. The comparison between theory
and experiment shows that, in high-mobility ${\it n}$-doped QWs,
the studied spin decoherence is controlled by electron-electron
collisions.
\end{abstract}

\maketitle

At present it is widely accepted that in zinc-blende-lattice {\it
n}-doped quantum well structures the spin-relaxation of conduction
electrons is dominated by the D'yakonov-Perel' (DP) mechanism. In
this mechanism the spin splitting of electronic states acts as an
effective magnetic field with the Larmor frequency ${\bm
\Omega}_{\bm k}$ dependent on the value and direction of the
electron wave vector ${\bm k}$. The related spin relaxation time
is given by \cite{dk}
\[
\frac{1}{\tau_s}\propto \langle \Omega^2 \tau \rangle,
\]
where $\tau$ is the microscopic relaxation time controlling spin
decoherence. Recently we have shown \cite{gi} that the inverse
time $\tau^{-1}$ has contributions not only from electron momentum
scattering processes responsible for the conductivity but also by
electron-electron collisions which {\it do not} affect the
electron mobility. Experimentally, the effect of electron-electron
collisions on the spin dynamics has been recently demonstrated by
Brand et al. \cite{harley1}. In Ref. \cite{gi} only the case of a
nondegenerate 2D electron gas is treated quantitatively while the
data \cite{harley1} are obtained on a degenerate high-mobility
gas. The purpose of this work is to extend the theory \cite{gi} to
arbitrary  electron degeneracy and compare the temperature
dependence of $\tau_s$ with the experiment \cite{harley1}.

A convenient form to represent temperature dependence of the spin
relaxation time is to write $\tau_s$ as
\begin{equation}\label{taush}
\tau_s^{-1}=\Omega_0^2\tau^*,
\end{equation}
where $\Omega_0$ is the effective Larmor frequency at the Fermi
energy at zero temperature and $\tau^*$ is a temperature-dependent
parameter which can be compared with the momentum relaxation time
$\tau_p$ obtained from measurement of the Hall mobility. The
representation (\ref{taush}) is usefully applied in the
scattering-dominated regime, where $\Omega_0\ll \tau^{-1}$,
realized at $T\sim 10$ K and higher. At the low temperature
$T=1.8$ K, the electron spin polarization evolves as heavily
damped oscillations of frequency $\Omega_0\approx0.19$ ps$^{-1}$
and $\tau^*$ is found from the exponential decay of these
oscillations. Thus, instead of $\tau_s$, we present in Fig.~1 the
temperature dependence of the above defined time $\tau^*$.

Values of $\tau^*$ extracted from the experiment are plotted in
Fig.1 by crosses together with the momentum relaxation time
$\tau_p$ (full circles). The spin polarization was monitored by
time-resolved optical response of a sample in which the 2D gas was
confined in a $(001)$-oriented $10$-nm {\it n}-doped GaAs/AlGaAs
quantum well structure. Electron concentration was estimated to be
$N_s=1.86\times10^{11}$ cm$^{-2}$ and Hall measurements showed
$N_s$ to be approximately constant at $T$ below $100$ K. The
transport relaxation time $\tau_p$ was extracted from the Hall
mobility.

\begin{figure}[t]
\epsfxsize=0.75\textwidth \epsfysize=0.85\textwidth \leavevmode
\centering{\epsfbox{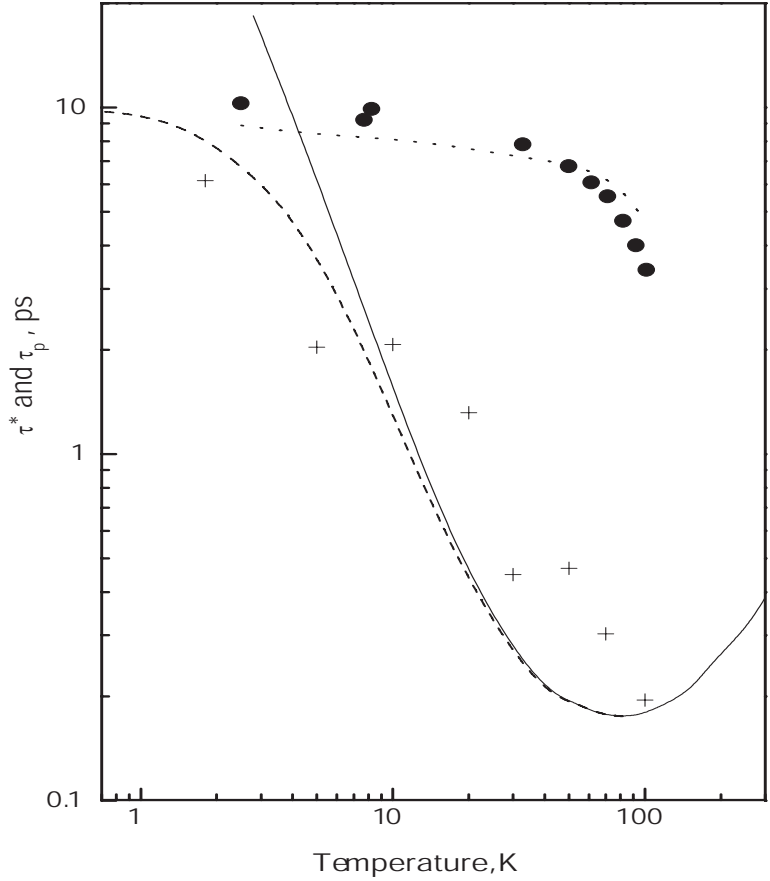}} \caption{{\small Temperature
dependencies of the scattering time $\tau^*$ controlling the
D'yakonov-Perel' spin relaxation, see Eq. (\ref{taush}), and the
transport scattering time $\tau_p$ which determines the mobility.
Experimental data are represented by crosses ($\tau^*$) and full
circles ($\tau_p$) \cite{harley1}. Theoretical curves for $\tau^*$
are calculated neglecting either electron-electron scattering
(dotted) or momentum scattering (solid) and taking into account
both scattering mechanisms (dashed).}}
\end{figure}

The inverse electron spin relaxation time in the symmetric GaAs QW
is given by \cite{book}
\begin{equation} \label{taus}
\frac{1}{\tau^s_{zz}} =\langle ({\Omega}^2_{1\bm k}+{
\Omega}^2_{3\bm k})\tau_p\rangle \:,
\end{equation}
where $\hbar\Omega_{1\bm k}=2\beta k(\langle k_z^2\rangle-
k^2/4)$, $\hbar\Omega_{3\bm k}=\beta k^3/2$, $\beta$ is the
constant describing spin-orbit splitting of the conduction band in
bulk GaAs and $\langle k_z^2\rangle$ is the quantum mechanical
average of the squared electron wave vector along the growth axis.
In Eq.~(\ref{taus}) electron-electron collisions are neglected.
The dotted line in Fig.~1 presents $\tau^*$ calculated from
(\ref{taus}) with $\tau_p$ given by the full circles. Definitely,
this line does not fit experimental points (crosses) and,
therefore, one can conclude that the account for momentum
scattering processes only is not sufficient to explain the
experimental results. The next step is to include
electron-electron collisions into consideration and calculate the
times $\tau_s$ and $\tau^*$ with allowance for both
electron-electron and momentum scattering.

In the frame of kinetic theory, the electron distribution in the
wave vector and spin spaces is described by a $2 \times 2$
spin-density matrix $ {\rho}_{\bm k} = f_{\bm k} + {\bm s}_{\bm k}
\cdot {\vec \sigma}$, where $\sigma_{\alpha}$ are the Pauli
matrices. Here $f_{\bm k} = \mbox{Tr}{ ({\rho}_{\bm k} /2)}$ is
the distribution function of electrons in the $\bm k$-space, and
${\bm s}_{\bm k}= \mbox{Tr}{[{\rho}_{\bm k} ({\vec \sigma}/2)]}$
is the average spin in the ${\bf k}$ state. If we neglect the spin
splitting then, for arbitrary degeneracy of an electron gas with
non-equilibrium spin-state occupation but equilibrium energy
distribution within each spin branch, the electron spin-density
matrix is the Fermi-Dirac distribution function with different
chemical potentials for the spin $\pm 1/2$. If the spin splitting
is non-zero but small as compared to $\hbar / \tau^*$, the
distribution function $\mbox{Tr}{[\rho_{\bm k}/2]}$ does not
change, whereas the spin vector acquires a correction $\delta {\bm
s}_{\bm k}$ proportional to the spin splitting.

The quantum kinetic equation for the spin pseudovector taking into
account both electron-electron collisions and elastic (or
quasi-elastic) momentum scattering has the form
\begin{equation} \label{kinetic}
\frac{\partial {\bm s}_{\bm k}}{\partial t} + {\bm \Omega}_{\bm k}
\times {\bm s}_{\bm k} + \frac{\delta{\bf s_k}}{\tau_p} + {\bm
Q}_{\bm k} \{ \delta{\bf s} , f\} = 0\:,
\end{equation}
where ${\bm Q}_{\bm k} \{ {\bm s}, f \}$ is the electron-electron
collision integral and $\tau_p$ is the momentum scattering time.
In the particular case of low spin polarization the
electron-electron scattering rate has the form \cite{gi_nato}
\begin{equation}\label{q_fd}
{\bf Q}_{\bf k} \{{\bf s}, f\} = \frac{2\pi}{\hbar} \sum_{\bm k',
\bm p, \bm p'} \delta_{\bm k+ \bm k', \bm p+ \bm p'}\: \delta (E_k
+ E_{k'} - E_p - E_{p'})
\end{equation}
\[
\times \left\{ 2 V_{\bm k - \bm p}^2 M({\bm k},{\bm k'},{\bm
p},{\bm p'})-V_{\bm k - \bm p}V_{\bm k - \bm p'} [M({\bm k},{\bm
k'},{\bm p},{\bm p'})+M({\bm k'},{\bm k},{\bm p},{\bm
p'})]\right\}.
\]
Here $V_{\bf q}$ is the Fourier transform of the electron-electron
interaction potential, $$M({\bf k},{\bf k'},{\bf p},{\bf p'})={\bm
s}_{\bm k} F({\bm k}';{\bm p},{\bm p}') - {\bm s}_{\bm p} F({\bm
p}';{\bm k},{\bm k}'),$$ $F({\bm k}_1;{\bm k}_2,{\bm k}_3)
=f_{{\bm k}_1} (1 - f_{{\bm k}_2} - f_{{\bm k}_3}) + f_{{\bm k}_2}
f_{{\bm k}_3}$ and $f_{\bm k}$ is the equilibrium Fermi-Dirac
distribution function. The term proportional to $2 V_{\bm k - \bm
p}^2$ is due to the direct Coulomb interaction whereas the term
proportional to $V_{\bm k - \bm p}V_{\bm k - \bm p'}$ comes from
the exchange interaction.

The spin-relaxation time governed by electron-electron collisions
was calculated allowing only for the linear-${\bm k}$ term in
${\bm \Omega}_{\bm k}$ and using a fixed value of $\tau_p$. We
used statically screened 2D Coulomb potential for $V_{\bm q}$. The
solid line in Fig.~1 shows temperature dependence of
$\tau^*=\tau_s^{-1}\Omega_0^{-2}$ calculated taking into account
electron-electron collisions only ($\tau_p=\infty$). A
non-monotonous behavior of this time can be understood as follows.
In the limit of low temperatures, for the degenerate
two-dimensional electron gas, $\langle\Omega_{\bf k}^2\rangle$
tends to a constant value $\Omega_0^2$ while the electron-electron
scattering rate $\tau_{ee}^{-1}$ vanishes as $T^2\ln{T}$.
Therefore, if momentum scattering is neglected the scattering time
$\tau^* \to \infty$. The allowance for electron momentum
scattering stabilizes both $\tau_s^{-1}$ and $\tau^*$ at $T=0$.
With rising temperature the role of electron-electron collisions
increases resulting in a decrease of $\tau^*$. In the opposite
limit of high temperatures the electron gas becomes nondegenerate
in which case $\tau_{ee}\propto T$, $\langle\Omega_{\bf
k}^2\rangle \propto T$ and the spin relaxation rate due to
electron-electron collisions increases with temperature according
the $T^2$ law \cite{gi}. The dependence $\tau^*(T)$ exhibits a
minimum near the transition from degenerate to nondegenerate
statistics when the chemical potential of electron gas reaches the
conduction band bottom.

Some discrepancy between the theoretical curve (dashed) and
experimental points (crosses) can probably be eliminated by taking
into account that, in the $10$-nm GaAs/AlGaAs QW, the electron
wave function has a quasi-2D character owing to its spread within
the well and electron tunnelling into the barriers \cite{glazov}.

In conclusion, we have shown both experimentally and theoretically
that the D'yakonov-Perel' spin relaxation may be controlled by
electron-electron collisions which do not affect the mobility in
the same way as by any other carrier scattering process.

This work is financially supported by the RFBR and by the
Programmes of Russian Ministry of Science and Presidium of RAS.
One of the authors (M.M.G.) is grateful to the Dynasty Foundation
and ICFPM for financial support.

\end{document}